\documentclass[
reprint,
 amsmath,amssymb,
 aps,
pra,
]{revtex4-2}

\usepackage{graphicx}
\usepackage{tikz}
\usepackage{dcolumn}
\usepackage{bm}
\usepackage{braket}

\begin{document}


\title{ Comment on J.Qin et al., Unconditional and   Robust Quantum Metrological Advantage beyond N00N States, PRL 130, 070801 (2023) and J.A. H.  Nielsen et.al., Deterministic Quantum Phase Estimation beyond N00N States. PRL 130,  123603 (2023).
 }

\author{Zden\v{e}k Hradil }
\affiliation{Department of Optics, Palack\' y University, 17.
listopadu 12,  779~00 Olomouc, Czech Republic}
\author{Jaroslav \v{R}eh\'{a}\v{c}ek}
\affiliation{Department of Optics, Palack\' y University, 17.
listopadu 12,  779~00 Olomouc, Czech Republic}

\email{hradil@optics.upol.cz}

\date{\today}


\begin{abstract}
\end{abstract}

\maketitle

The papers by Qin \emph{et al.}~\cite{Qin2023} and Nielsen \emph{et al.}~\cite{Nielsen_23}
report an unconditional quantum metrological advantage based on squeezed-vacuum phase sensing.
The principal evidence for this claim is the Quantum Fisher Information (QFI) evaluated
\emph{per trial} or \emph{per detected photon}, which is subsequently interpreted as a measure
of metrological performance.

This interpretation is statistically unjustified because it does not account for the total
experimental resources required to construct an estimator.
Consequently, the reported protocols do not provide an unconditional metrological advantage
over classical phase-sensing strategies when resource accounting is performed consistently.

Quantum metrology is an estimation problem rather than a single-measurement problem.
The performance of any estimation protocol is determined by an estimator constructed from a
finite data set and is therefore governed by the Cram\'er--Rao inequality
\begin{align}
(\Delta\phi)^2 \ge \frac{1}{nF},
\label{CR}
\end{align}
where $F$ denotes the Fisher information associated with a single detection event and
$n$ is the number of statistically independent repetitions used to construct the estimator. 

Though the experiments reported in
Refs.~\cite{Qin2023,Nielsen_23}
employ different physical realizations, leading to a factor-of-two difference in the definition of the Fisher information and the mean photon number owing to the use of two-mode and single-mode squeezed states, respectively, this distinction does not affect the present argument. For simplicity, we therefore restrict our discussion to the single-mode squeezed vacuum, for which
\begin{align}
F=2\sinh^2(2r)=8\bar n(\bar n+1),
\end{align}
where
\[
\bar n=\sinh^2 r
\]
is the mean photon number.

The crucial point is that the experimental resource is \emph{not} the photon number per trial
but the total number of photons required to complete the estimation protocol,
\begin{align}
N=n\bar n .
\end{align}
{\em Consequently, the QFI per trial is not an operational measure of metrological
performance.} Omitting the repetition number $n$ removes an essential part of the resource
accounting and inevitably leads to misleading conclusions regarding scaling and possible quantum advantage.  
The omission of $n$ is not a minor technical simplification. It fundamentally changes the
statistical meaning of the Cram\'er--Rao bound. A single measurement cannot define an
estimation precision; only an estimator constructed from repeated observations possesses
a well-defined variance.

This distinction becomes particularly important for squeezed-vacuum protocols.  Although the
QFI per trial increases quadratically with the mean photon number, the number
of repetitions required to reach the asymptotic Gaussian regime increases whenever the
likelihood exhibits multiple narrow extrema. Once this statistically necessary repetition
number is taken into account, the total resource scaling becomes classical as $1/N$. Quantum resources
may improve the numerical prefactor, but not this asymptotic scaling.

The central role of repeated observations is already evident from the reported experiments.
Nielsen \emph{et al.}~\cite{Nielsen_23} explicitly evaluate phase uncertainties from
datasets containing $M=1000$ statistically independent homodyne measurements for a mean photon number of only $\bar n=1.8$.
In contrast, Qin \emph{et al.}~\cite{Qin2023} extract the phase information from calibration curves whose statistical fluctuations are smaller than the symbol size in Fig.~3(a), likewise implying very large datasets.
In both cases, the phase uncertainty is determined by the combined effect of repeated observations and Fisher information through the product $MF$, whereas the claimed ``quantum advantage'' is inferred solely from the QFI per trial.

The comparison with ideal NOON states suffers from the same conceptual problems.
NOON states exhibit $N$-fold phase ambiguities and therefore cannot achieve a phase variance
proportional to $1/N^2$ without additional prior information identifying the correct interference
maximum. Such prior information itself constitutes an experimental resource. Ignoring this resource
while quoting Heisenberg scaling produces an operationally meaningless benchmark.

 But there is further conceptual difficulty, which  
 makes those phase detection  schemes impractical, namely strong dependence of the performance on the detected phase. The reported  "quantum advantage" is obtained only in a narrow phase interval where the Fisher information reaches its maximum. Away from this operating point the sensitivity rapidly deteriorates, as demonstrated by the experimentally measured phase variance itself. A metrological protocol whose performance depends critically on the unknown value of the parameter cannot be regarded as an unconditional estimation strategy without including the resources required for adaptive localization of the operating point.

An elementary analogy illustrates the same conceptual problem. A stopped clock shows the exact time twice a day. Its indication is therefore perfectly accurate at those two particular instants, but this does not make it a useful clock: without knowing when those instants occur, the accuracy of the indication cannot be exploited for time determination. Likewise, a metrological protocol may exhibit exceptionally high resolution at particular values of the unknown parameter while providing little information about where those values occur.

Our criticism concerns exclusively the statistical interpretation of the reported results and not the experimental implementation itself.
Unfortunately, these misconceptions are no longer confined  only just to the two papers discussed here.
Similar claims continue to appear in recent high-profile publications
\cite{Peng2026},
indicating that the distinction between ultimate statistical limits and operationally
achievable resolution remains insufficiently appreciated.

QFI has become a widely used surrogate for metrological performance. It is not. QFI may even diverge as the detected signal vanishes, while the achievable resolution remains strictly limited \cite{Hradil_19}. Quantum metrology is ultimately concerned not with preparing quantum states, but with inferring unknown parameters from finite experimental data. Quantum physics generates the data; statistics determines the resolution. Maximizing QFI by itself is neither a sufficient nor a necessary condition for maximizing operational resolution. The widespread practice of equating high QFI with quantum metrological advantage, particularly when accompanied by incomplete resource accounting, should therefore be reconsidered. More broadly, the methodology by which quantum metrological advantage is defined and demonstrated deserves renewed scrutiny. Operational quantum metrology requires consistent resource accounting, explicit estimator construction, and statistical validation~\cite{Hradil2026}.

\section*{Acknowledgment}
We acknowledge the support of the   Czech  Science Foundation   under the grant agreement 26-22242J.

\bibliography{MasterBib}

\end{document}